\documentclass[twocolumn,showpacs,preprintnumbers,amsmath,amssymb,showkeys]{revtex4}
\usepackage{graphicx}% Include figure files
\usepackage{dcolumn}% Align table columns on decimal point
\usepackage{bm}% bold math
\begin{document}
\draft
\date{\today}
\title{Some remarks on R\'{e}nyi relative entropy in a thermostatistical framework}
\author{G. B. Ba\u{g}c\i} \thanks{Corresponding Author}
\email{gbb0002@unt.edu}
\address {Department of Physics, University of North Texas, P.O. Box 311427, Denton, TX 76203-1427,
USA}

\pagenumbering{arabic}

\begin{abstract}
In ordinary Boltzmann-Gibbs thermostatistics, the relative entropy
expression plays the role of generalized free energy, providing the
difference between the off-equilibrium and equilibrium free energy
terms associated with Boltzmann-Gibbs entropy. In this context, we
studied whether this physical meaning can be given to R\'{e}nyi
relative entropy definition found in the literature from a
generalized thermostatistical point of view. We find that this is
possible only in the limit as $q$ approaches to 1. This shows that
R\'{e}nyi relative entropy has a physical (thermostatistical)
meaning only when the system can already be explained by ordinary
Boltzmann-Gibbs thermostatistics. Moreover, this can be taken as an
indication of R\'{e}nyi entropy being an equilibrium entropy since
any relative entropy definition is a two-probability generalization
of the associated entropy definition. We also note that this result
is independent of the internal energy constraint employed. Finally,
we comment on the lack of foundation of R\'{e}nyi relative entropy
as far as its minimization (which is equivalent to the maximization
of R\'{e}nyi entropy) is considered in order to obtain a stationary
equilibrium distribution since R\'{e}nyi relative entropy does not
conform to Shore-Johnson axioms.

\end{abstract}

  \narrowtext
\newpage \setcounter{page}{1}

\maketitle

\section{\protect\bigskip Introduction}

\bigskip Recently there has been a growing interest in generalized
entropies such as Tsallis [1], R\'{e}nyi [2] and Sharma-Mittal [3]
entropies in the context of a generalized thermostatistics. Although
R\'{e}nyi entropy has been introduced by A. R\'{e}nyi as early as
1961, the applications of this form of entropy have been for example
in the fields of quantum computation [4], information theory [5, 6]
and chaotic one-dimensional maps [7]. Only recently, some authors
have studied R\'{e}nyi entropy in the framework of a generalized
thermostatistics [8-12]. Lenzi et al. [8] and Bashkirov [11] for
example showed that it results in a power-law equilibrium
distribution by maximization whereas Parvan and Bir\'{o} [12]
concluded that R\'{e}nyi entropy also satisfies zeroth law of
thermodynamics. In this Letter, we will focus on the physical
meaning of R\'{e}nyi relative entropy (also called divergence or
cross-entropy) in the framework of a generalized thermostatistics.
Before proceeding further, it is important to assess the importance
of relative entropy in entropy maximization related issues and in
thermostatistics in general.For this purpose, let us write
Boltzmann-Gibbs (BG) entropy which reads

\begin{equation}
S_{BG}(p)=-\sum\limits_{i}^{W}p_{i}\ln p_{i},
\end{equation}

where p$_{\text{i }}$ is the probability of the system in the ith
microstate, W is the total number of the configurations of the
system. Note that Boltzmann constant k is set to unity throughout
the paper. The corresponding relative entropy is called
Kullback-Leibler entropy (K-L) [13] and it is given by

\begin{equation}
K[p\Vert q]\equiv \sum_{i}p_{i}\ln (p_{i}/q_{i}).
\end{equation}

Note that it is a convex function of ${p}_{i}$, always non-negative
and equal to zero if and only if $p=q$. The probabilities
${q}_{i}$'s are called prior or reference distributions. K-L entropy
can be thought as a generalization of BG entropy in the sense that
both are equal to one another, apart from a multiplicative constant,
when the prior distribution in relative entropy definition is known
with certainty i.e., a probability of one is assigned to it.
Therefore, it is always possible to obtain BG entropy as a
particular case of corresponding relative entropy expression, so
called K-L entropy. The converse is not true since K-L relative
entropy is a two-probability distribution generalization of BG
entropy. This situation can be compared to the case of R\'{e}nyi and
BG entropies: R\'{e}nyi entropy is considered to be a generalization
of BG entropy simply due to the fact that its parameter can be
adjusted in such a way that it results in BG entropy as a particular
case. Whenever the R\'{e}nyi index $q$ becomes 1, we obtain BG
entropy as a particular case. In this sense, any relative entropy
definition associated with a particular entropy, be it R\'{e}nyi or
Tsallis entropies, is a generalization of that particular entropy in
terms of probabilities whereas generalized entropies such as
R\'{e}nyi or Tsallis entropies are seen to be generalization in
terms of some parameter $q$ although the nature of this parameter is
not the same in these aforementioned cases. Second issue regarding
the importance of the concept of relative entropy is that ordinary
BG entropy cannot be generalized to continuum rigorously just by
changing summation to integration since it fails to be invariant
under different parametrizations. Moreover, it will not be bounded
neither from below nor above (see Ref.[14] and references therein).
Contrary to these problems with ordinary BG entropy in its
generalization to the continuum case, the relative entropy
definition does not face any of these problems. Therefore, relative
entropy is more general in its domain of applicability since it can
be used in the continuum case unlike ordinary BG entropy. All of the
remarks above can be summarized by the statement that the concept of
relative entropy is a generalization of the corresponding entropy
definition both in terms of probability distributions and continuum
case. It is our aim in this paper to present some results related to
the definition of R\'{e}nyi relative entropy concerning its physical
meaning and relation to internal energy constraints. But since
R\'{e}nyi relative entropy is understood to be a generalization of
R\'{e}nyi entropy, what can be said about R\'{e}nyi relative entropy
has important bearings on R\'{e}nyi entropy itself. In addition to
these important features of relative entropy in a thermostatistical
framework, it is also an important measure of complexity whose uses
range from the numerical analysis of protein sequences [15], pricing
models in the market [16], to medical decision making [17]. The
outline of the Letter is as follows: In the next Section, we revisit
the physical meaning of the ordinary relative entropy in the case of
BG entropy. We then study R\'{e}nyi relative entropy in a
thermostatistical framework in Section III. Section IV will be
devoted to the study of the axiomatic foundation of R\'{e}nyi
relative entropy. The conclusions will be discussed in Section V.

\section{\protect\bigskip Physical meaning of Kullback-Leibler
relative entropy}

\bigskip In order to study the physical meaning of any relative entropy in a
thermostatistical framework, one has first to obtain the equilibrium
distribution associated with the entropy of that particular
thermostatistics. In this Section, we will maximize BG entropy
subject to some constraints following the well known recipe of
entropy maximization. Let us assume that the internal energy
function is given by $U=\sum\limits_{i}\varepsilon _{i}p_{i}$, where
$\varepsilon_{i}$ denotes the energy of the ith microstate. In order
to get the equilibrium distribution associated with BG entropy, we
maximize the following functional

\begin{equation}
\Phi (p)=-\sum\limits_{i}^{W}p_{i}\ln p_{i}-\alpha
\sum\limits_{i}^{W}p_{i}-\beta
\sum\limits_{i}^{W}\varepsilon_{i}p_{i},
\end{equation}

where $\alpha$ and $\beta$ are Lagrange multipliers related to
normalization and internal energy constraints respectively. Equating
the derivative of the functional to zero, we obtain

\begin{equation}
\frac{\delta \Phi (p)}{\delta p_{i}}=-\ln \widetilde{p}_{i}-1-\alpha
-\beta \varepsilon_{i}=0.
\end{equation}

Tilde denotes the equilibrium distribution obtained by the
maximization of BG entropy. By multiplying Eq. (4) by
$\widetilde{p}_{i}$ and summing over i, using the normalization and
internal energy constraints, we have

\begin{equation}
\alpha +1=\widetilde{S}_{BG}-\beta \widetilde{U}.
\end{equation}

Substitution of Eq. (5) into Eq. (4) results in the following
equilibrium distribution

\bigskip
\begin{equation}
\widetilde{p}_{i}=e^{-\widetilde{S}_{BG}}e^{\beta
\widetilde{U}}e^{-\beta \varepsilon_{i}}.
\end{equation}

If we now use the equilibrium distribution $\widetilde{p}$ as the
reference distribution in K-L entropy, we can write

\begin{equation}
K[p\Vert \widetilde{p}]=\sum_{i}p_{i}\ln (p_{i}/\widetilde{p}_{i}).
\end{equation}

The equation above can be rewritten as

\begin{equation}
K[p\Vert \widetilde{p}]= -S_{BG}-\sum_{i}p_{i}\ln \widetilde{p}_{i}.
\end{equation}

We then insert the equilibrium distribution given by Eq. (6) in the
equation above to find

\begin{equation}
K[p\Vert \widetilde{p}]=
-S_{BG}-\sum_{i}p_{i}(-\widetilde{S}_{BG}+\beta \widetilde{U}-\beta
\varepsilon_{i}).
\end{equation}

Carrying out the summation, we have

\begin{equation}
K[p\Vert \widetilde{p}]= -S_{BG}+\widetilde{S}_{BG}-\beta
\widetilde{U}+\beta U,
\end{equation}

which can be cast into the form

\begin{equation}
K[p\Vert \widetilde{p}]= \beta (F_{BG}-\widetilde{F}_{BG}).
\end{equation}

The free energy term is given as usual by $F=U-S_{BG}/\beta$. The
result above shows us that the physical meaning of the K-L entropy
is nothing but the difference of the off-equilibrium and equilibrium
free energies when the reference distribution is taken to be the
equilibrium distribution given by Eq. (6) above.  This result can be
used, for example, to study equilibrium fluctuations or
non-equilibrium relaxation of polymer chains [18].

\bigskip
\bigskip
\bigskip
\bigskip

\section{R\'{e}nyi relative
entropy as a generalized free Energy}

After studying the physical meaning of K-L entropy in the previous
Section, we are now ready to study the meaning of R\'{e}nyi relative
entropy in generalized thermostatistical framework. In order to do
this, we begin by writing R\'{e}nyi entropy [2]
\begin{equation}
S_{R}(p)=\frac{1}{1-q}\ln (\sum\limits_{i}p_{i}^{q}),
\end{equation}
where the parameter $q$ is an arbitrary real number. R\'{e}nyi
entropy is equal to or greater than zero for all values of the
parameter $q$ and concave for $q\leq1$. It reduces to BG entropy
given by Eq. (1) as the parameter $q$ approaches 1. Using internal
energy constraint in terms of escort probabilities i.e.,
$U_{q}=\frac{\sum\limits_{i}\varepsilon
_{i}p_{i}^{q}}{\sum\limits_{j}p_{j}^{q}}$, the functional to be
maximized reads

\begin{equation}
\Phi _{R}(p)=\frac{1}{1-q}\ln (\sum\limits_{i}^{W}p_{i}^{q})-\alpha
\sum\limits_{i}^{W}p_{i}-\beta \frac{\sum\limits_{i}^{W}\varepsilon_{i}p_{i}^{q}}{%
\sum\limits_{i}^{W}p_{j}^{q}}.
\end{equation}

We take the derivative of this functional and equate it to zero in
order to obtain the following

\begin{equation}
\frac{\delta \Phi _{R}(p)}{\delta p_{i}}=\frac{q}{1-q}\frac{\widetilde{p}%
_{i}^{q-1}}{\sum\limits_{j}\widetilde{p}_{j}^{q}}-\alpha -\beta ^{\ast }q%
\widetilde{p}^{q-1}(\varepsilon_{i}-\widetilde{U}_{q})=0,
\end{equation}

where $\beta^{\ast}$ is given by

\begin{equation}
\beta ^{\ast }=\frac{\beta }{\sum\limits_{j}\widetilde{p}_{j}^{q}}.
\end{equation}

Multiplying the equation above by $\widetilde{p}_{i}$ and summing
over the index i, we find

\begin{equation}
\alpha =\frac{q}{1-q}.
\end{equation}

Note that tilde denotes that the distribution is calculated at
equilibrium. Substituting this explicit expression of $\alpha$ into
Eq. (14), we calculate the explicit form of equilibrium distribution
. It reads

\begin{equation}
\widetilde{p}_{i}=(\frac{1}{e^{(1-q)\widetilde{S}_{R}}}-(1-q)\beta
^{\ast }(\varepsilon _{i}-\widetilde{U}_{q}))^{1/(1-q)}.
\end{equation}

The R\'{e}nyi relative entropy [19] reads

\begin{equation}
I_{q}[p\Vert r]=\frac{1}{q-1}\ln
(\sum\limits_{i}p_{i}^{q}r_{i}^{1-q}).
\end{equation}

This definition of R\'{e}nyi relative entropy too is always
non-negative and equal to zero if and only if $p=r$. It also reduces
to K-L entropy as the parameter $q$ approaches 1. We then substitute
equilibrium distribution in Eq. (17) into the relative entropy
definition above and obtain

\begin{equation}
I_{q}[p\Vert \widetilde{p}]=\frac{1}{q-1}\ln (\sum\limits_{i}p_{i}^{q}((%
\frac{1}{e^{(1-q)\widetilde{S}_{R}}}-(1-q)\beta ^{\ast }(\varepsilon _{i}-%
\widetilde{U}_{q})))).
\end{equation}

Having summed up over indices, we obtain

\begin{equation}
I_{q}[p\Vert \widetilde{p}]=\frac{1}{q-1}\ln (e^{(1-q)(S_{R}-\widetilde{S}%
_{R})}-(1-q)\beta ^{\ast \ast }(U_{q}-\widetilde{U}_{q})),
\end{equation}

where $\beta ^{\ast \ast }$ is given by

\begin{equation}
\beta ^{\ast \ast }=\frac{\beta }{\sum\limits_{j}\widetilde{p}_{j}^{q}}%
\sum\limits_{i}p_{i}^{q}.
\end{equation}

Inspection of Eq. (20) shows that one cannot cast it into the form
of free energy differences associated with R\'{e}nyi-related
quantities due to the logarithmic term involved. Indeed, one needs
to apply Taylor expansion two times, first to the exponential term
in the parentheses and second to the logarithmic term itself. Having
made these two Taylor series expansions about $q=1$, we finally
arrive a familiar result i.e.,

\begin{equation}
I_{q}[p\Vert \widetilde{p}]=\beta(F_{BG}-\widetilde{F}_{BG}),
\end{equation}

where free energy expressions are exactly the same as in the BG
case. This result is trivial and equal to the expression obtained in
Section II by using BG entropy and K-L entropy. It should be noted
that the first Taylor expansion turned the R\'{e}nyi entropy into BG
entropy while second Taylor expansion turned the Lagrange multiplier
and internal energy functions into their corresponding BG values.

It is important to underline one crucial point: we have maximized
R\'{e}nyi entropy with escort distribution and used this equilibrium
distribution as the reference distribution for the associated
relative entropy expression. However, if we try to maximize it with
ordinary constraint, then we obtain

\begin{equation}
\widetilde{p}_{i}=[(1-\beta \frac{q-1}{q}(\varepsilon _{i}-\widetilde{U}_{q}%
))\sum\limits_{j}\widetilde{p}_{j}^{q}]^{1/(q-1)}.
\end{equation}

for the equilibrium distribution. It is obvious that the
substitution of Eq. (23) into relative entropy expression given by
Eq. (18) does not yield to a result which can be written as
difference of free energies for all $q$ values. Again, the relative
entropy will be a generalized free energy only in the limit as $q$
approaches 1. This shows that the choice of internal energy
constraint does not matter at all in assessing the physical meaning
of R\'{e}nyi relative entropy.

\section{Axiomatic foundation of R\'{e}nyi relative entropy}

The stationary equilibrium distribution associated with a particular
entropy expression can be found either by the maximization of that
entropy or the minimization of the corresponding relative entropy.
In this sense, it is expected both of these methods to be consistent
and well-founded in terms of axioms. Shore and Johnson considered
the axioms, which must be satisfied by any relative entropy
expression if they were expected to lead to minimum relative entropy
for the stationary equilibrium distribution in a consistent manner
[20, 21]. These axioms are given by

\begin{enumerate}
  \item Axiom of Uniqueness: If the same problem is solved twice,
  then the same answer is expected to result both times.
  \item Axiom of Invariance: The same answer is expected when the
  same problem is solved in two different coordinate systems, in
  which the posteriors in the two systems should be related by the
  coordinate transformation.
  \item Axiom of System Independence: It should not matter whether
  one accounts for independent information about independent systems
  separately in terms of their marginal distributions or in terms of
  the joint distribution.
  \item Axiom of Subset Independence: It should not matter whether
  one treats independent subsets of the states of the systems in
  terms of their separate conditional distributions or in terms of
  the joint distribution.
  \item Axiom of Expansibility: In the absence of new information,
  the prior should not be changed.
\end{enumerate}

These authors were able to show that the satisfaction of all these
axioms resulted in a relative entropy expression consistent with the
principle of minimum relative entropy and could be summarized in one
simple expression: according to Shore and Johnson, any relative
entropy $J[p\Vert r]$ with the prior $r_{i}$ and posterior $p_{i}$
which satisfies five very general axioms, must be of the form

\begin{equation}
J[p\Vert r]=\sum\limits_{i}p_{i}h(p_{i}/r_{i}),
\end{equation}

for some function $h(x)$. Ordinary relative entropy i.e., K-L
entropy is in accordance with Shore-Johnson theorem since the
function $h(x)$ can be identified as the natural logarithm. In the
case of R\'{e}nyi relative entropy given by Eq. (18), we see that it
cannot be cast into a form which will conform to the Shore-Johnson
theorem for any function $h(x)$. In other words, there is no
axiomatic basis for the use of R\'{e}nyi relative entropy in order
to get the stationary equilibrium distribution for a possible
generalized thermostatistics.

\section{Conclusions}

The relative entropy acts as a generalized free energy in the
ordinary thermostatistical framework when one makes use of the
associated equilibrium distribution as the reference distribution.
In this Letter, we investigated whether R\'{e}nyi relative entropy
can play the role of a generalized free energy in a
thermostatistical framework. We found that this is possible only in
the limit as $q$ approaches to 1. This shows that R\'{e}nyi relative
entropy has a thermostatistical meaning only when the system is in a
state of ordinary BG thermostatistics. This can be taken as an
indication of R\'{e}nyi entropy being an equilibrium entropy and
nothing but an approximation to ordinary BG thermostatistics since
any relative entropy definition is a two-probability generalization
of the associated entropy definition. The choice of internal energy
constraint too does not solve the problem. Still, relative entropy
acts as a generalized free energy only in the $q$=1 limit. This is
different than the case of BG or Tsallis entropy [21]. In Tsallis
case in particular, one does not need to make any approximation,
which in turn makes the physical interpretation valid for all
positive q values. This is reminiscent of the fact that R\'{e}nyi
entropy provides the same answer at the microcanonical structure as
BG entropy, yielding $lnW$ [22]. As R\'{e}nyi entropy yields the
same result as BG entropy in the microcanonical case, so does
R\'{e}nyi relative entropy in its connection to free energies. In
other words, it is redundant to use R\'{e}nyi entropy in the
microcanonical case and it is so to use R\'{e}nyi relative entropies
in calculating nonequilibrium fluctuations of polymer chains since
we already have K-L entropy for this purpose [18]. These results
regarding the physical meaning of R\'{e}nyi relative entropy are
similar to the results derived from Lesche stability condition [23]
for R\'{e}nyi entropy since both considers deformation of
probabilities and concludes R\'{e}nyi entropy makes sense only in
the limit when $q$ approaches 1 in which case it becomes BG entropy
(corresponding relative entropy becomes K-L entropy). Lastly, it can
be noted that the results presented in this paper are also supported
by the original methodology of orthodes devised by Boltzmann when
applied to R\'{e}nyi entropy since it leads one to deduce that
R\'{e}nyi entropy is an equilibrium entropy [24]. It is interesting
to note that this result has also been reached independently in Ref.
[25] by using a different method than the method of orthodes
employed in Ref. [24] thereby making this view even stronger. Our
approach to this issue explains one more difficulty arising from the
comparison of the works of Refs. [24] and [25] since in the former
escort distribution has been employed whereas ordinary constraint
has been used in the latter. In our view, it is not surprising that
the same conclusion has been reached concerning R\'{e}nyi entropy
being an equilibrium entropy since associated relative entropy
possesses a physical meaning in both cases only when the parameter
$q$ approaches 1, thereby making the difference in employed
constraints redundant. Apart from the verification of this result,
it can be noted that the novelty here is the understanding of this
entropy to be an approximation to the ordinary BG entropy in the
thermostatistical framework in a generalized setting of the
definition of corresponding relative entropy. Lastly, we have shown
that R\'{e}nyi relative entropy does not have a sound axiomatic
foundation since it does not conform to Shore-Johnson axioms.
Therefore, any attempt to minimize R\'{e}nyi relative entropy in
order to obtain an associated stationary equilibrium distribution in
the context of generalized thermostatistics fails to be consistent.

\end{document}